# Short-range magnetic order and temperature-dependent properties of cupric oxide

X. Rocquefelte,[1] M.-H. Whangbo,[2] A. Villesuzanne,[3] S. Jobic,[1] F. Tran,[4] K. Schwarz[4] and P. Blaha[4]

[1] Institut des Matériaux Jean Rouxel, Université de Nantes, CNRS, UMR6502, Boîte Postale 32229, 44322 NANTES Cedex 3, France

[2] Department of Chemistry, North Carolina State University, Raleigh, North Carolina 27695-8204

[3] Institut de Chimie de la Matière Condensée de Bordeaux, UPR9048, ICMCB-CNRS, av. Dr. A. Schweitzer, 33608 Pessac Cedex, France

[4] Institute of Materials Chemistry, Vienna University of Technology, Getreidemarkt 9/165-TC, A-1060 Vienna, Austria

Email: Xavier.Rocquefelte@cnrs-imn.fr

**Abstract**

The temperature dependence of the optical and magnetic properties of CuO were examined by means of hybrid density functional theory calculations. Our work shows that the spin exchange interactions in CuO are neither fully one-dimensional nor fully three-dimensional. The large temperature dependence of the optical band gap and the $^{63}$Cu nuclear quadrupole resonance frequency of CuO originate from the combined effect of a strong coupling between the spin order and the electronic structure and the progressive appearance of short-range order with temperature.

In strongly-correlated transition-metal oxides the coupling between spin, charge, orbital and vibrational degrees of freedom favours the emergence of exotic properties such as colossal magnetoresistance, high-temperature superconductivity or multiferroicity. Cupric oxide CuO has been extensively studied since the discovery of cuprate superconductors due to its close resemblance in structure and magnetic properties. In particular, some features of cuprate superconductors appear also in CuO, for example, Zhang-Rice singlets at the top of its valence band [1], charge stripes [2] and a strong magneto-lattice coupling [3]. Recently, CuO has been found to be multiferroic at high temperature [4].

The optical band gap and the absorption edge of CuO exhibit a surprisingly strong dependence on temperature as shown by Marabelli *et al.* [5]. They concluded that the electronic states involved in the optical absorption processes have a localized character, originating from a strong electron-lattice coupling. Above $T_N$ = 230 K, the three-dimensional (3D) magnetic ordering temperature, CuO is usually described as a quasi one-dimensional (1D) Heisenberg antiferromagnet with a strong spin exchange (*i.e.*, J = -67 ± 20 meV) [6 – 9]. Thus, one may wonder whether the strong temperature dependence of the optical band gap and the absorption edge in CuO arise from a strong coupling of the electronic structure with the spin order rather than with the lattice. In this Letter we show that this is indeed the case by examining the spin exchange interactions, the dielectric function, the band gap and the electric field gradients (EFGs) of CuO on the basis of first principles density functional theory (DFT) [10] electronic structure calculations.

The structure of CuO can be viewed as built upon corner- and edge-sharing square-planar $CuO_4$ units, which form $(-Cu-O-)_\infty$ zigzag chains running along the [101] and [10$\bar{1}$] directions of the chemical unit cell (a, b, c), *i.e.* the a′ and c′ directions of the magnetic cell, respectively (see fig.1). The $Cu^{2+}$ spin moments order to have an incommensurate antiferromagnetic (AFM) structure with propagation vector q = (0.506, 0, -0.483) below $T_N$ = 230 K, and subsequently a commensurate structure with q = (0.5, 0, -0.5) below $T_L$ = 213 K [6 – 8], in which $Cu^{2+}$ spins have AFM and ferromagnetic (FM) arrangements along the [10$\bar{1}$] and [101] directions, respectively [7]. The dimensionality of the magnetic structure has been controversial; the magnetic, thermodynamic and neutron data of CuO above $T_N$ can be reproduced using either a 1D AFM chain or a two-dimensional (2D) AFM lattice model. This picture of low-dimensional magnetism is supported by spin dimer analysis based on extended Hückel tight binding (EHTB) calculations [11,12], according to which CuO can be described as 1D AFM chains. However, in a recent pseudopotential self-interaction-free (pseudo-SIC) DFT study, Filippetti and Fiorentini [13] reported that CuO is an antiferromagnet with fully 3D spin exchange interactions, and the magnetic orbital of each square-planar $CuO_4$ unit is not the Cu $d_{x^2-y^2}$ but the Cu $d_{z^2}$ orbital. The latter conclusions are inconsistent with the observed low-dimensional magnetic properties of CuO and the well-established fact that the highest-lying d-block orbital of a square-planar complex of a transition-metal element is not represented by the $d_{z^2}$ but by the $d_{x^2-y^2}$ orbital [12,14]. Therefore, an accurate and independent determination of the spin exchange interactions is highly desirable.

To evaluate the spin exchange interaction of CuO, we initially considered the nine ordered spin states (*i.e.*, $AF_x$, $AF_y$, $AF_z$, $AF_a$, $AF_{-101}$, FM, $FM_1$, $FM_x$ and $FM_z$) depicted in fig. 2 assuming collinear arrangements of magnetic moments. These ordered spin states can be constructed by using the magnetic unit cell consisting of eight formula units (FUs) shown in fig. 1.

DFT calculations were performed for these states on the basis of the room-temperature (RT) single-crystal X-ray structure [15, 16] using the WIEN2k program package [18]. The present calculations employed a hybrid functional based on the PBE [19] functional of the generalized gradient approximation (GGA). In hybrid functionals [20], the exchange energy is obtained by mixing Hartree-Fock (HF) and DFT exchange energies in certain proportion. The optimal fraction α of the Hartree-Fock exchange varies according to the systems under examination (nature of the elements, local arrangement, and dimensionality), the property, and the exchange-correlation functionals employed in the mixing [21 – 23]. The implementation of this approach in solid state DFT codes is quite recent, but the relevance in describing transition-metal oxide [23 – 25] and excited state properties (*e.g.*, optical band gap, spin exchange, etc.) has been demonstrated in several studies [24 – 27]. In particular, the solutions provided by such hybrid calculations go beyond the Kohn-Sham ground-state solutions, allowing reproducing quite accurately time-consuming many-body calculations [26]. In addition, it should be noticed that the present results for excited states properties (magnetism, optics) are corroborated by ground-state property results like the electric-field gradient (EFG), which was shown to be extremely sensitive to the treatment of the correlated electrons, as shown in ref. [28]. To ensure the validity of the present hybrid generalized gradient approximation (hybrid/GGA) calculations, and more specifically the value of the mixing parameter, α = 0.15, we have considered the low-temperature AFM structure of CuO. Particularly, our calculations lead to the EFG of the Cu atom of $-8.50 \times 10^{21}$ V/m$^2$ (see below) and the spin moments on the Cu and O atoms of 0.65 and 0.12 $\mu_B$, respectively, which compare well with the experimental values of - $7.80\times10^{21}$ V/m$^2$, 0.65 and 0.14 $\mu_B$ [7] respectively. The theoretical value for the optical band gap of about 1.42 eV compares also well with the experimental value of 1.4 eV [29]. While our results are overall in reasonable agreement with previous DFT calculations, their values for the band gap and Cu spin moments of 1.9 eV and 0.74 $\mu_B$ in LDA+U ($U_{eff}$ = 6.5 eV) [30] and of 2.2 eV and 0.72 $\mu_B$ in pseudo-SIC approaches [13] were significantly overestimated.

The relative energies of the nine ordered spin states obtained from our hybrid/GGA calculations are summarized in table I. To extract the spin exchange parameters $J_a$, $J_b$, $J_x$, $J_z$ and $J_2$ of CuO as defined in figs. 1 and 2, we express the total spin exchange interactions of the nine ordered spin states in terms of the spin Hamiltonian $\hat{H} = -\sum_{i<j} J_{ij}\hat{S}_i \cdot \hat{S}_j$, where $J_{ij}$ (=$J_a$, $J_b$, $J_x$, $J_z$ or $J_2$) refers to the spin exchange parameter for the spin sites i and j [31]. With this convention, an AFM spin exchange interaction is represented by a negative J value. Thus by mapping the energy differences between the ordered spin states given by the spin Hamiltonian onto those given by hybrid/GGA calculations, we obtain the values of the five spin exchange parameters listed in table II.

In agreement with the neutron diffraction study [7] our calculations show that the spin arrangement AF$_z$ is the most stable magnetic state. In the study of Filippetti and Fiorentini [13] all spin exchange interactions are comparable in magnitude (*i.e.*, $J_z \approx 2J_x \approx -3J_b \approx 3J_2 \approx -5J_a$), and only $J_z$ and $J_2$ favor AFM interactions. However, the present hybrid/GGA calculations (with $\alpha$ = 0.15) show that only three interactions are significant ($J_z \approx 4J_2 \approx 7J_a$) and favor AFM interactions. $J_z$ is the strongest spin exchange interaction, as already found in the EHTB study ($J_z \approx 16 J_x \approx 19 J_b \approx 19 J_2 \approx 310 J_a$) [11]. Table I shows that only three terms ($J_z$, $J_2$ and $J_x$) contribute to the stability of the AF$_z$ state. $J_x$ was previously considered as the driving force of the FM couplings along x [13], but is very weak. Thus, the present calculations evidence that the FM layers of the $AF_z$ state are a direct consequence of the fact that the 1D AFM chains (Cu1-Cu2 and Cu5-Cu6 in fig.1) made up of the dominant superexchange interaction $J_z$ are also antiferromagnetically coupled through the Cu-O…O-Cu super-superexchange interaction $J_2$ (Cu2-Cu5 and Cu1-Cu6).

Given the experimental estimate of J = -67±20 meV [6 – 9], the exchange parameters obtained from the hybrid/GGA calculations with $\alpha$ = 0.15 are overestimated (*i.e.*, $J_z$ = -128.8 meV, table II). These small interaction energies are very sensitive to the chosen DFT functional. As expected, if we use a larger mixing value of $\alpha$ = 0.25 (PBE0 hybrid functional) [21,22], we obtain smaller J values in much better agreement with the experimental estimation. In particular, the strongest interaction $J_z$ = -80.5 meV, is then in good agreement with the reported experimental value. However the larger fraction of HF exchange leads to a significant overestimation of the optical band gap of about 2.4 eV, the EFG of about $-11.96 \times 10^{21}$ V/m$^2$ and the Cu spin moment of 0.74 $\mu_B$, while the O spin moment is then underestimated, 0.09 $\mu_B$. As a matter of fact, with a larger mixing value the Cu-d states are more localized leading to a reduced orbital overlap with O-p states. Consequently the ionicity of the Cu-O chemical bond is increased, leading to a relocalization of the magnetic moment on Cu site, an increase of the band gap, and a reduction of the spin exchange mediated by the orbital overlap between Cu and O atoms.

We should keep in mind, that the choice of the mixing parameter in such mean-field approaches is a matter of compromise: high values for spin exchange estimation, while smaller values for spectroscopic data simulation. In any case, the hybrid/GGA calculations with both $\alpha = 0.15$ and 0.25 give the same trends in the spin exchange parameters, namely, $J_z/J_2 \approx 4$ and other J values are much smaller. These findings are in strong disagreement with previous calculations by Filippetti *et al.* [13] and restore the previously accepted view of magnetic interactions in CuO [6–9,11]. The above picture of CuO, deduced from our hybrid/GGA calculations, predicts that the spin correlation lengths along the c′- and a′-directions slightly above $T_N$ should be controlled by $J_z$ and $J_2$, respectively. Consistent with this prediction, the observed spin correlation length along the c′-direction is about 3.5 times greater than that along the a′-direction (*i.e.*, 700 vs. 200 Å) [32], which is consistent with the fact that $J_z/J_2 \approx 4$.

From the viewpoint of the calculated spin exchange parameters, the temperature dependence of the magnetic susceptibility of CuO, which shows a broad maximum at around $T_{max} = 550$ K [6-9], can be described as follows: at high temperature (around $T_{max}$ and above), CuO exhibits short-range 1D AFM order based on $J_z$. As the temperature is lowered, CuO undergoes short-range 2D AFM order based on $J_z$ and $J_2$ (above $T_N$ and below $T_{max}$), which is followed by incommensurate 3D AFM long-range order based largely on $J_z$, $J_2$ and $J_a$ (below $T_N$ and above $T_L$) and then by commensurate 3D AFM long-range order (below $T_L$). As can be seen from fig. 1, both $J_a$ and $J_b$ do not contribute to the stability of the $AF_z$ phase but can introduce spin frustration between adjacent 2D rectangular nets defined by $J_z$ and $J_2$. The latter might be responsible for the incommensurate 3D AFM long-range order between 213 - 230 K, in which every second 1D chains made up of $J_z$ undergoes a spiral spin order responsible for the ferroelectric polarization [4]. The occurrence of short-range magnetic order is consistent with the observations that no paramagnetic scattering is detected down to 550 K in the paramagnetic regime (T > $T_N$) [7], and that more than 70% of the spins are ordered above $T_N$ according to the specific-heat measurements [33].

Let us now explore the temperature dependence of the optical properties of CuO. By considering low- and high-temperature crystal structures of CuO [16], we note that the small changes in the crystal structure with temperature has a negligible effect on the calculated dielectric function ε, and then on the temperature dependence of the optical band gap of CuO [5]. Thus, electron-lattice coupling cannot explain the unusual temperature dependency. However, the temperature-induced rearrangement of the spin order, described above, may influence the electronic structure and its related optical transitions. Upon raising the temperature above $T_N$, the thermal energy induces spin flips, which destroy the 3D long-range AFM order, but short range 2D/1D AFM order remains. When the 1D AFM order is destroyed by spin flipping, chain segments with FM spin order occur along the c′ direction, which explains the continuous increase in the magnetic susceptibility in the temperature region up to $T_{max}$ = 550 K. The electronic structures for various short-range spin ordered states of CuO, which occur when the temperature is raised, were simulated by hybrid/GGA calculations for CuO with a (4a × b × 4c) supercell containing 64 FUs. It is convenient to label the short-range spin ordered states of CuO in terms of the percentage, P, of FM segments present within the AFM chains of $Cu^{2+}$ ions along the c′-direction. Thus, P = 0 % and P = 100 % describe the $AF_z$ and FM states, respectively. Fig. 3 compares the imaginary parts of the dielectric function $ε_2$ (related to the optical absorption) calculated for the states with P = 0, 6.25, 12.5, 25, 31.25, 37.5 %. Note that in our model P is expected to increase with increasing temperature, but of course never reaches 100 % (a hypothetical ordered FM state). The predicted optical response corresponding to a hypothetical FM state is also given in fig. 4.

As can be seen from figs. 3 and 4, the RT experimental $\varepsilon_2$ curve [34, 35] shows two broad absorption peaks. The latter are not reproduced by the electronic structure of either the $AF_z$ or the FM state. However, the $\varepsilon_2$ curves calculated for the electronic states with short-range spin order show an optical response intermediate between those of the $AF_z$ and FM states, hence reproducing the essential feature of the experimental $\varepsilon_2$ curve. Note that the lower-energy absorption is overestimated by the $AF_z$ state while it is absent in the FM state. Thus, fig. 3 reveals that, with increasing temperature (*i.e.*, with increasing the size of FM segments within each chain of $J_z$), the lower-energy absorption peak undergoes three significant changes; a reduction in the intensity, a displacement toward the lower energy and a decrease in the steepness of the absorption edge. The latter two findings explain the large temperature dependence (from 10 K to 300 K) of the experimentally observed absorption edge [5], namely, a gradual decrease in the band gap by about 0.25 eV and in the steepness of the absorption edge. Our simulations show that the optical band gap is reduced by about 0.3 eV when P increases from 0 % to 37.5 %. The present study clearly evidences that the strong temperature dependence of the optical properties of CuO is a consequence of strong coupling between the spin order and the electronic structure, and not a strong electron-lattice coupling as previously proposed [5].

To better understand the origin of the strong coupling between the optical response and the degree of spin order in CuO, we compare the density of states (DOS) calculated for the P = 0, 37.5 and 100 % states (fig. 4). In all cases the absorption edge is mainly due to a charge transfer transition from the O 2p to the empty Cu $3d_{x^2-y^2}$ bands. Both the majority and the minority spins are involved in this transition for the P = 0 and 37.5 % states, but only the minority spins contribute for the P = 100 % state. The latter explains why the low-energy absorption is absent in the FM state. The DOS plots show that the O 2p and Cu 3d states are strongly affected on going from the $AF_z$ state to the P = 37.5 % state and then to FM state (P = 100 %). The $3d_{x^2-y^2}$ bands progressively increase in width and hence induce a reduction in the optical gap when the temperature is raised (fig. 4). It should be emphasized that although the experimental $\varepsilon_2$ data appear featureless, they were obtained in two independent studies [34, 35]. The broad feature of $\varepsilon_2$ at RT is not a consequence of experimental limitation, but reflects a broadened DOS at RT that is brought about by the introduction of FM domains within each chain of $J_z$.

The essence of the above observations can be readily accounted for by considering the electronic structure of a spin-1/2 uniform AFM chain made up of the spin exchange $J_z$. If this chain is described by the nearest-neighbour hopping integral t and the on-site repulsion U, the up- and down-spin bands have a width proportional to t in the FM state, but $t^2/U$ in the AFM state. For a strongly AFM chain, t >> $t^2/U$, so that the bandwidth is significantly narrower in the AFM state than in the FM state, as has been pointed out by Xiang *et al*. in their analysis of the dependence of the electronic structure of $SrFeO_2$ on its spin arrangement [36]. As the temperature is raised, spin flips take place in the chain hence generating locally FM domains in the AFM chain of $J_z$. The FM domains generate discrete energy levels, the high- and the low-lying levels of which would lie above and below the narrow band associated with the AFM regions of the chain, respectively. This explains why the loss of 3D AFM long-range order in CuO decreases the band gap and the steepness of the absorption edge and why the states participating in the excitation have a localized character.

Finally, we show that the temperature dependence of the $^{63}$Cu NQR frequency in CuO is another signature of strong coupling between the spin order and the electronic structure in CuO. The $^{63}$Cu NQR frequency of CuO increases linearly with temperature [37, 38], and this finding is not explained by considering the lattice expansion with increasing temperature. The $^{63}$Cu NQR frequency is proportional to the EFG and is very sensitive to small changes in the electronic structure. The strong coupling between the spin order and the electronic structure discussed above should affect the electron density around the Cu atoms in CuO, and hence the electric field gradient (EFG) of the Cu atoms. We calculated the EFGs of the Cu atoms by performing hybrid/GGA calculations with $\alpha = 0.15$ for three states, namely the low-temperature 3D long-range AF$_z$ spin order (P = 0%), the hypothetical 3D long-range FM spin order (P = 100%) and the 2D/1D short-range spin order state, P = 37.5%, which leads to the best agreement with the RT experimental optical data. Our calculations show that the EFGs of the Cu atoms increase almost linearly with increasing P. For example, the calculated EFG is $-8.50 \times 10^{21}$ V/m$^2$ for the AF$_z$ state (P = 0 %), $-8.78 \times 10^{21}$ V/m$^2$ for the P = 37.5 % state, and $-9.28 \times 10^{21}$ V/m$^2$ for the FM state (P = 100 %). Thus, our calculations predict that, on going from the low-temperature (P = 0%) to the high-temperature (P = 37.5%) magnetic structure, the EFG should increase by about 3 %. This is consistent with the experimental observation between 77 K and 295 K (EFG = $-7.80 \times 10^{21}$ and $-8.04 \times 10^{21}$ V/m$^2$, respectively, by using Q = $-0.211$ barn for $^{63}$Cu).

In summary, the temperature-induced change in the dimensionality of the magnetic structure of CuO from 1D to 3D is well explained in terms of the spin exchange parameters deduced from the present work. The temperature dependence of the magnetic susceptibility, optical band gap and $^{63}$Cu NQR frequency in CuO originates from a strong coupling between the spin order and the electronic structure. It remains to explore if the present theoretical scenario could be generalized to other highly correlated transition-metal oxides with strong J values, such as the High-Tc cuprate parent compounds.

The authors thank Centre de Calcul Intensif des Pays de la Loire for generous computing resources. The work at NCSU was supported by the Office of Basic Energy Sciences, Division of Materials Sciences, U. S. Department of Energy, under Grant DE-FG02-86ER45259. PB and FT were supported by the Austrian Science Fund P20271-N17.

Table I. Energy expressions per Cu atom and the corresponding relative energies[a] (in meV) deduced from hybrid/GGA calculations ($\alpha$ is the fraction of Hartree-Fock exchange).

| State | Energy expression | $\alpha = 0.15$ | $\alpha = 0.25$ |
| --- | --- | --- | --- |
| $AF_z$ | $(-J_x + J_z + J_2)/4$ | -40.4 | -25.8 |
| $AF_{-101}$ | $(-J_a + 2J_b + J_x + J_z - J_2)/4$ | -17.4 | -11.7 |
| $AF_a$ | $+J_a/8$ | -2.3 | -0.5 |
| $FM_x$ | $-J_x/4$ | -0.3 | -0.7 |
| $FM_1$ | $(-J_a - 2J_b - J_x - J_z - J_2)/8$ | 20.7 | 11.7 |
| $AF_x$ | $(+J_x - J_z + J_2)/4$ | 25.3 | 16.0 |
| $FM_z$ | $-J_z/4$ | 32.0 | 20.1 |
| $AF_y$ | $(+J_a + 2J_b - J_x - J_z - J_2)/4$ | 36.6 | 25.0 |
| $FM$ | $(-J_a - 2J_b - J_x - J_z - J_2)/4$ | 41.5 | 23.5 |

[a] The reference state, with respect to which the relative energies are given, is defined as the hypothetical state for which the total spin exchange energy is zero. For example, $E(\text{reference}) \equiv [2E(FM_1) - E(FM)]/2$.

Table II. Geometrical and spin exchange J (in meV) parameters of CuO obtained from the hybrid/GGA calculations with $\alpha$ = 0.15 and $\alpha$ = 0.25

|  | $J_a$ | $J_b$ | $J_x$ | $J_z$ | $J_2$ [31] |
|---|---|---|---|---|---|
| $d_{Cu-Cu}$ (Å) [15] | 2.901 | 3.083 | 3.173 | 3.749 | 5.129 |
| ∠Cu-O-Cu (°)[15] | 95.71 | 104.02 | 108.91 | 145.81 |  |
| Calculated J with $\alpha$ = 0.15 | −18.2 | +4.2 | +2.6 | −128.8 | −30.1 |
| Calculated J with $\alpha$ = 0.25 | −4.0 | +3.5 | +3.0 | −80.5 | −19.6 |

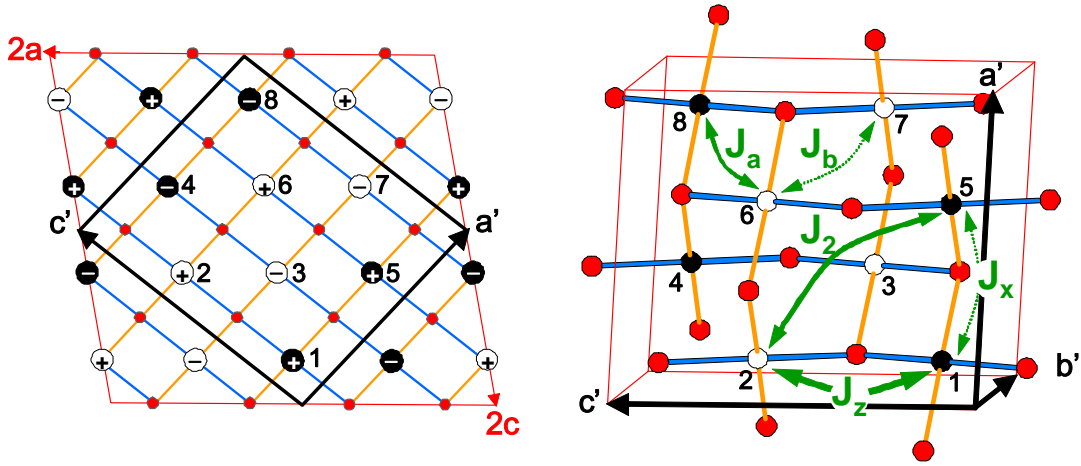

Fig. 1: Schematic representations of the magnetic structure of CuO observed below $T_L = 213$ K. The left panel shows a projection along the b axis of the $2a \times 2c$ supercell containing 16 $Cu^{2+}$ ions. The magnetic unit cell (a′, b′, c′), shown in black, contains eight $Cu^{2+}$ ions. The plus and minus signs indicate the position of copper atoms along the b′ axis. The right panel shows a 3D perspective view of the magnetic structure. The O atoms are presented by red circles and the $Cu^{2+}$ sites along the b-direction are depicted as filled and empty circles representing up-spin and down-spin, respectively. The definitions of the five spin exchange paths $J_z$, $J_2$, $J_a$, $J_b$ and $J_x$ in the magnetic unit cell (a′, b′, c′) are also given.

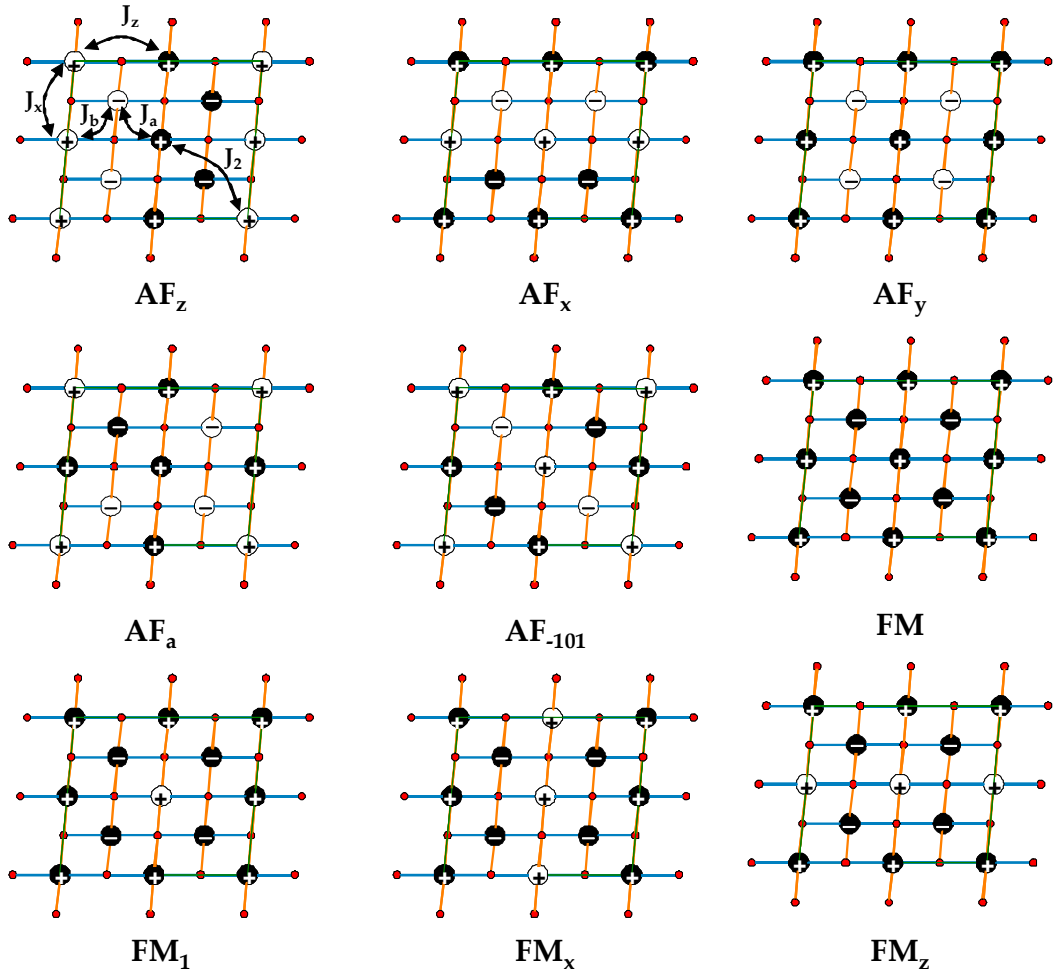

Fig. 2. Schematic projection views, along the b′ axis, of the nine ordered spin arrangements of the Cu$^{2+}$ ions in CuO.

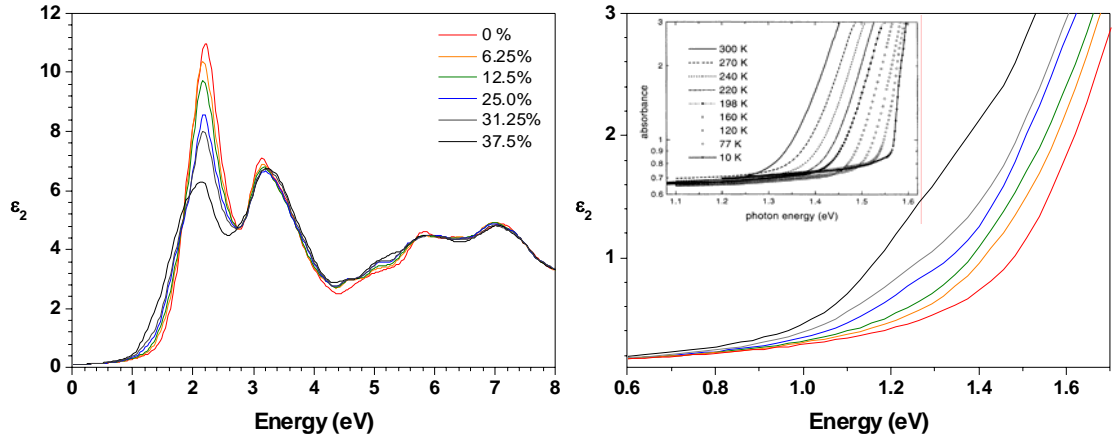

Fig. 3: Isotropic imaginary part ε₂ of the dielectric function of CuO as a function of the incident light energy. The left panel shows the ε₂ curves obtained from the hybrid/GGA calculations with α = 0.15 for the magnetic states of CuO with P = 0, 6.25, 12.5, 25.0, 31.25 and 37.5 %. The right panel shows a zoomed-in view around the absorption edge, where the inset shows the experimental absorbance of CuO at the absorption edge as a function of temperature [5].

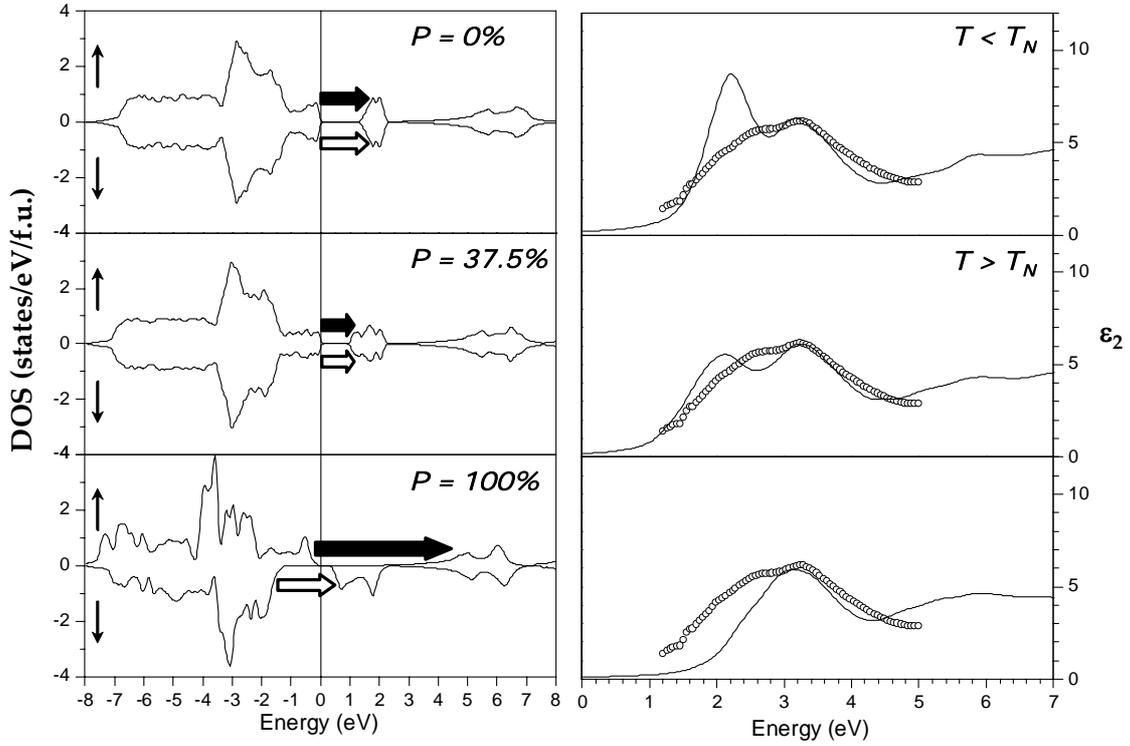

Fig. 4: DOS plots (left panel) and $\varepsilon_2$ curves (right panel) obtained from the hybrid/GGA calculations with $\alpha = 0.15$ for the P = 0, 37.5 and 100 % states of CuO. In the right panel, the experimental $\varepsilon_2$ plot obtained at room temperature is given as empty circles [34].